\documentclass[sigconf]{acmart}

\copyrightyear{2022}
\acmYear{2022}
\setcopyright{acmcopyright}
\acmConference[CIKM '22] {Proceedings of the 31st ACM International Conference on Information and Knowledge Management}{October 17--21, 2022}{Atlanta, GA, USA.}
\acmBooktitle{Proceedings of the 31st ACM International Conference on Information and Knowledge Management (CIKM '22), October 17--21, 2022, Atlanta, GA, USA}
\acmPrice{15.00}
\acmISBN{978-1-4503-9236-5/22/10}
\acmDOI{10.1145/3511808.3557398}

\usepackage{multirow}
\graphicspath{ {./figures/} }

\begin{document}

\title{Modeling Dynamic Heterogeneous Graph and Node Importance for Future Citation Prediction}

\author{Hao Geng}
\affiliation{%
  \institution{SKLSDE, School of Computer Science at Beihang University}
  \city{Beijing}
  \country{China}
  \postcode{100191}
}
\email{genghao@buaa.edu.cn}

\author{Deqing Wang}
\authornote{Corresponding author.}
\affiliation{%
  \institution{SKLSDE, School of Computer Science at Beihang University}
  \city{Beijing}
  \country{China}
  \postcode{100191}
}
\email{dqwang@buaa.edu.cn}

\author{Fuzhen Zhuang}
\affiliation{%
  \institution{SKLSDE, School of Computer Science \& Institute of Artificial Intelligence at Beihang University}
  \city{Beijing}
  \country{China}
  \postcode{100191}
}
\email{zhuangfuzhen@buaa.edu.cn}

\author{Xuehua Ming}
\affiliation{%
  \institution{SKLSDE, School of Computer Science at Beihang University}
  \city{Beijing}
  \country{China}
  \postcode{100191}
}
\email{xhming@buaa.edu.cn}

\author{Chenguang Du}
\affiliation{%
  \institution{SKLSDE, School of Computer Science at Beihang University}
  \city{Beijing}
  \country{China}
  \postcode{100191}
}
\email{duchenguang@buaa.edu.cn}

\author{Ting Jiang}
\affiliation{%
  \institution{SKLSDE, School of Computer Science at Beihang University}
  \city{Beijing}
  \country{China}
  \postcode{100191}
}
\email{royokong@buaa.edu.cn}

\author{Haolong Guo}
\affiliation{%
  \institution{SKLSDE, School of Computer Science at Beihang University}
  \city{Beijing}
  \country{China}
  \postcode{100191}
}
\email{ghl_123@buaa.edu.cn}

\author{Rui Liu}
\affiliation{%
  \institution{SKLSDE, School of Computer Science at Beihang University}
  \city{Beijing}
  \country{China}
  \postcode{100191}
}
\email{lr@buaa.edu.cn}

\renewcommand{\shortauthors}{Hao Geng et al.}

\begin{abstract}

Accurate citation count prediction of newly published papers could help editors and readers rapidly figure out the influential papers in the future. Though many approaches are proposed to predict a paper's future citation, most ignore the dynamic heterogeneous graph structure or node importance in academic networks. To cope with this problem, we propose a Dynamic heterogeneous Graph and Node Importance network (DGNI) learning framework, which fully leverages the dynamic heterogeneous graph and node importance information to predict future citation trends of newly published papers. First, a dynamic heterogeneous network embedding module is provided to capture the dynamic evolutionary trends of the whole academic network. 
Then, a node importance embedding module is proposed to capture the global consistency relationship to figure out each paper's node importance. Finally, the dynamic evolutionary trend embeddings and node importance embeddings calculated above are combined to jointly predict the future citation counts of each paper, by a log-normal distribution model according to multi-faced paper node representations. Extensive experiments on two large-scale datasets demonstrate that our model significantly improves all indicators compared to the SOTA models.

\end{abstract}

\ccsdesc[500]{Information systems~Data mining}

\keywords{citation count prediction, dynamic heterogeneous graph, node importance estimation}

\maketitle

\section{Introduction}

Predicting the impact of research papers is of great significance for science researchers to find out the most promising research topic to study and identify significant works from a sea of scientific literature \cite{2016-CanScientificImpactBePredicted}. As there is no precise definition of the impact of a scientific research, citation counts of scientific papers are usually taken as the estimation \cite{2009-OpenAccessAndGlobalParticipationInScience, 2011-CitationCountPrediction, 2016-QuantifyingTheEvolutionOfIndividualScientificImpact}.

However, the task of predicting citation counts is challenging and nontrivial, due to the following reasons. Firstly, as existing publications own citation counts in each year, when a new publication emerging, there does not exist any historical citations, leading to the lack of label information to train the model. Secondly, in the heterogeneous academic network, each paper is associated with heterogeneous information such as authors, venues and fields, and how to make full use of these heterogeneous information turns to be a great challenge. Thirdly, all nodes in the academic network keep continuously evolving with changeable states, making it crucial to properly capture the dynamics. To follow previous footsteps, we summarize the existing methods for citation count prediction task in the following two categories.

The first category predicts early published papers' citation counts with previous citations available \cite{2019-PredictingCitationCountsBasedOnDeepNeuralNetworkLearningTechniques, 2017-OnPredictivePatentValuation, 2014-ModelingAndPredictingPopularityDynamicsViaReinforcedPoissonProcesses, 2013-QuantifyingLongtermScientificImpact, 2016-OnModelingAndPredictingIndividualPaperCitationCountOverTime, 2018-ModelingAndPredictingCitationCountViaRecurrentNeuralNetworkWithLSTM}. One part of these methods design parametric patterns to model the paper citation trend, including the log-normal intensity function in Wang et al. \cite{2013-QuantifyingLongtermScientificImpact}, the reinforced Poisson process in Shen et al. \cite{2014-ModelingAndPredictingPopularityDynamicsViaReinforcedPoissonProcesses} and the recency-weighted effect in Liu et al. \cite{2017-OnPredictivePatentValuation}. Other methods utilize neural networks to capture the temporal patterns in historical citations, such as the Recurrent Neural Network (RNN) in Yuan et al. \cite{2018-ModelingAndPredictingCitationCountViaRecurrentNeuralNetworkWithLSTM} and the seq2seq framework in Abrishami \cite{2019-PredictingCitationCountsBasedOnDeepNeuralNetworkLearningTechniques}. However, there methods are unable to predict citation counts for newly published papers, since there exists no historical citations.

The second category predicts citation counts for newly published papers without historical citations. As a paper carries multidimensional information including its related authors, keywords, reference papers and venues, some researchers extract hand-crafted features to represent a paper \cite{2007-EstimatingNumberOfCitationsUsingAuthorReputation, 2016-CanScientificImpactBePredicted, 2012-ToBetterStandOnTheShoulderOfGiants, 2011-CitationCountPrediction, 2014-CitationImpactPredictionForScientificPapersUsingStepwiseRegressionAnalysis}. For instance, Dong et al. \cite{2016-CanScientificImpactBePredicted} represents a paper by 6 types of factors, including author, reference, topic, venue, social and temporal features. Yan et al. \cite{2011-CitationCountPrediction} extracts rank-based features such as author rank and venue rank, as a part of paper features. However, these feature engineering methods require expert knowledge and manual labour for feature designing, which is time-consuming and cannot utilize the power of heterogeneous and evolving characteristics of nodes. HINTS \cite{2021-HINTS} is the first work as we known to design an end-to-end framework for predicting new paper citation counts without feature engineering. Specifically, HINTS takes the whole academic network as a sequence of heterogeneous graphs, and combines the temporally aligned Graph Neural Network (GNN), the Recurrent Neural Network (RNN) and a time series generator to learn representations of the sequence. Although creative and rational, it fails to make full use of heterogeneous academic network features and the significance or popularity of a node in the graph, and thus causing the prediction accuracy lower. 

In this paper, we propose a new framework based on \textbf{D}ynamic Heterogeneous \textbf{G}raph and \textbf{N}ode \textbf{I}mportance Network, named \textbf{DGNI}, which models the dynamic evolutionary trends of the academic network and each paper's node importance on the global scale to predict future citation counts of newly-published papers. DGNI is divided into three parts: dynamic heterogeneous network embedding module, node importance embedding module, and time series generation module. 

Firstly, in dynamic heterogeneous network embedding module, we use Heterogeneous Graph Neural Network (HGNN) on snapshots of the academic network from different timestamps, together with RNN-based model to jointly model time series features. So the module can capture the dynamic evolutionary trends of the whole academic network before the publication of the papers. 

Since the heterogeneous graph neural network can only capture the local consistency relationship of academic network, neglecting the significance of the global consistency relationship. In node importance embedding module, we propose a node importance embedding module to calculate each paper's node importance on the global scale to capture the global consistency relationship. This is consistent with our intuition that a paper with higher importance and influence in the academic community tends to receive more citations in the subsequent years. 

Finally, in time series generation module, the dynamic evolutionary trends embeddings and node importance embeddings calculated above are transformed into the parameters of the time series generation module, using a simple multilayer perceptron (MLP). Following \cite{2013-QuantifyingLongtermScientificImpact, 2021-HINTS}, we use a log-normal distribution model to generate the prediction citation counts sequence for each newly published paper.

In summary, our main contributions can be summarized as follows:
\begin{itemize}
  \item We solve the challenging cold start problem in time series citation prediction. To be specific, it refers to the prediction of citation count for newly published articles without historical citation count values.

  \item We propose a novel framework named DGNI for citation time series prediction, which leverages both the local consistency relationship and global consistency relationship of the heterogeneous academic network. Hence our model can make full use of the dynamic academic network and node importance to predict future citation count of newly published papers.

  \item We conduct extensive experiments on two large-scale real-world academic network datasets, and the experimental results illustrate that our model outperforms the SOTA models by 11.39\% improvement in terms of MAE and 12.90\% improvement in terms of RMSE.
\end{itemize}

The rest of this paper is organized as follows: Section 2 introduces necessary definitions and makes a formal definition of the problem we tackle. Section 3 introduces the motivation and framework of our proposed model DGNI, and further elaborate each component of our model. Section 4 evaluates the performance of DGNI by experiments and analyses. Section 5 reviews the related works of citation count prediction, node importance estimation and heterogeneous graph representation learning. Section 6 makes a conclusion to the entire paper.

\section{Preliminaries}

In this section, we introduce necessary definitions used in the paper and make a formal definition of the problem we study.

\subsection{Definitions}

\subsubsection{Heterogeneous Academic Network} 

A heterogeneous academic network is a special kind of heterogeneous information network (HIN), which consists of multiple types of nodes and edges to represent academic networks. It can be defined as a graph $\mathcal{G=(V, E)}$ with node mapping function $\varphi: \mathcal{V\to T}$ and edge mapping function $\phi: \mathcal{E\to R}$ where $\mathcal{|T|+|R|>}$2.

The types of nodes in heterogeneous academic network include paper, venue, field and author. It is widely noticed that the papers are the central nodes and other nodes are neighbors. The types of edge include publish (paper-venue), write (author-paper), contain (paper-field) and cite (paper-paper).

\subsubsection{Metapath and metapath-based subgraph}  Metapath is defined as a path with the following form: $A_1 \xrightarrow{R_1} A_2 \xrightarrow{R_2} \cdots \xrightarrow{R_{l-1}} A_l$ (abbreviated as $A_1 A_2 \cdots A_l$), where $A_i \in \mathcal{V}, R_i \in \mathcal{E}$. The metapath describes a composite relation between node types $A_1$ and $A_l$, which expresses specific semantics.

Given a metapath $\phi_p$ of a heterogeneous graph $\mathcal{G}$, the metapath-based subgraph of graph $\mathcal{G}$is defined as a graph composed of all neighbor pairs based on metapath $\phi_p$.

\subsubsection{Dynamic Heterogeneous Network} 

A dynamic heterogeneous network is a sequence of heterogeneous academic network from 1 to T year: $\langle \mathcal{G}_t\rangle_{t=1}^T=\{\mathcal{G}^1,\mathcal{G}^2,...,\mathcal{G}^T\}$, where $\mathcal{G}^t=(\mathcal{V}^t, \mathcal{E}^t)  (1\leq t\leq T) $ is the heterogeneous academic network in \emph{t} th year.

\subsubsection{Node importance} 

A node importance $s \in \mathbb{R}^+$ is a non-negative real number representing the significance or the popularity of an entity in a knowledge graph. For instance, the gross of the movie or the voting number for the movie on the website can be regarded as the node importance in movie knowledge graphs. The specific importance value of a node is collected from the real scenarios and obtained after the log transformation.

\begin{figure*}[htbp]
  \centering
  \centering
  \includegraphics[width=0.8\textwidth]{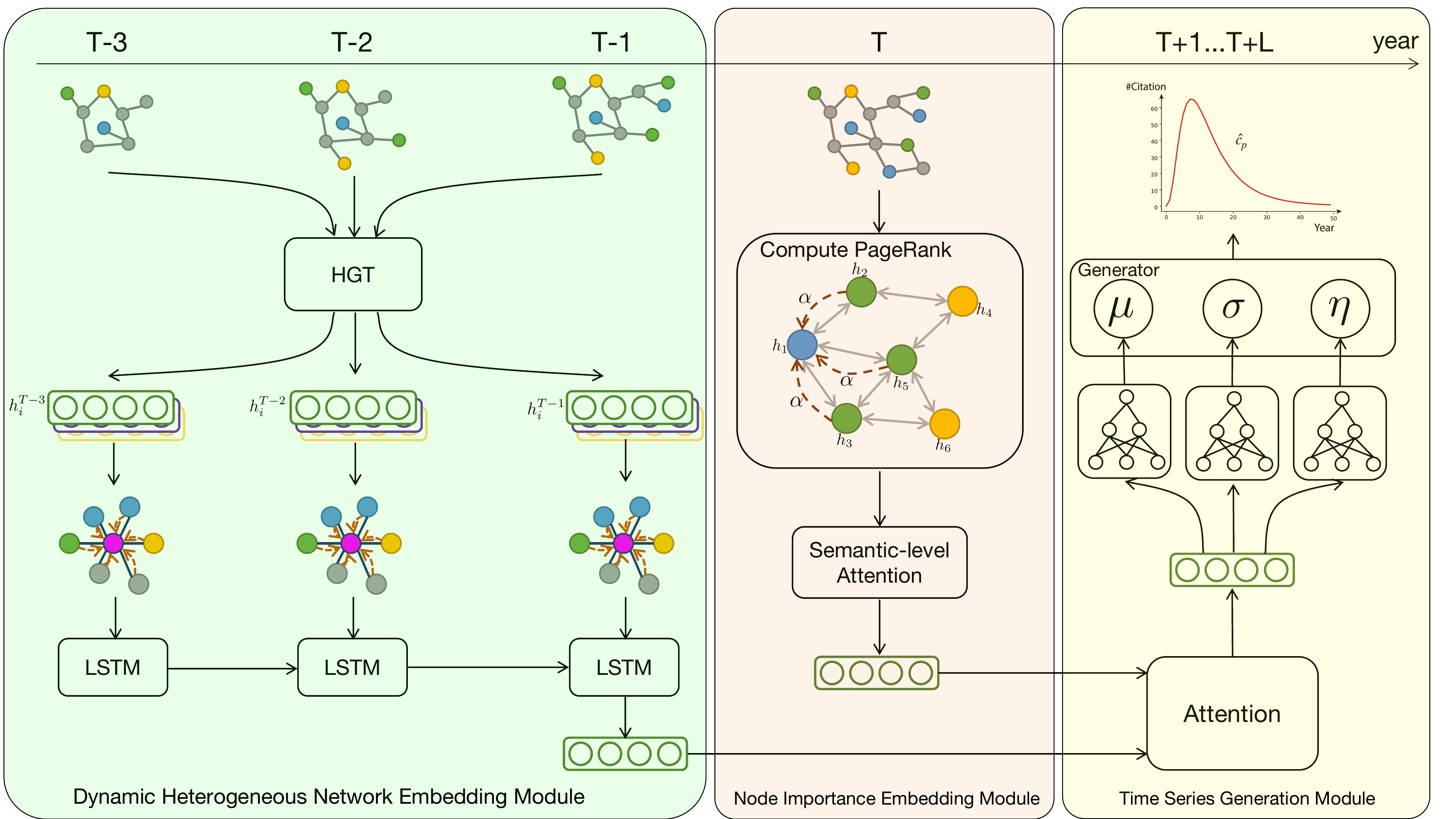}
  \vspace{-0.2cm}
  \caption{The overall architecture of our proposed model DGNI. To predict the citation count time series of a new paper $p$ published in year $T$, DGNI first learns the heterogeneous network features from years before $T$ to generate the fake embeddings of $p$ before its publication. Then we compute the node importance of $p$ by our proposed node importance module. After that, the above embeddings are fed into the final time series generation module to generate the cumulative citation of $p$ in $1$-$L$ years after its publication.}
  \vspace{-0.3cm}
  \label{fig:framework}
\end{figure*}

\subsection{Problem Formalization} 

Given a dynamic heterogeneous network $\langle \mathcal{G}_t\rangle_{t=1}^T$ and a target paper \emph{p}, paper citation time series prediction aims to learn a function \emph{f}: $(\langle \mathcal{G}_t\rangle_{t=1}^{T}, p) \to \{c_p^{T+1}, c_p^{T+2},...,c_p^{T+L}\}$ that predicts the citation time series of the target paper \emph{p} in the following \emph{L} years after the publication year \emph{T}.

\section{Methodology}

In this section, we introduce our proposed model DGNI for citation count prediction. First, we describe our motivation to design the architecture of DGNI, then we present the framework of DGNI, as shown in Fig. \ref{fig:framework}. Next, we elaborate the details of three components in DGNI: dynamic heterogeneous network embedding module, node importance embedding module, and time series generation module.

\subsection{Framework of DGNI}

The key idea of our model is to learn a continuously evolving vector representation of each node from the snapshots of the academic network in different periods, so the node representation can reflect the node's evolutionary trend. Powered by such representations, the dynamics of the academic network can be well captured, making it easier to predict future citations for new papers. As shown in Fig. \ref{fig:framework}, our proposed method DGNI is composed of three modules: dynamic heterogeneous network embedding module, node importance embedding module, and time series generation module. 

According to \cite{2021-HINTS}, the impact of paper can be predicted by modeling changes of snapshots of dynamic heterogeneous networks in different periods. In our dynamic heterogeneous network embedding module, instead of simply using RGCN \cite{2018-RGCN} to capture the heterogeneity of the academic network, which keeps distinct non-sharing weights for node types and edge types alone and is insufficient to capture heterogeneous properties, we use the SOTA heterogeneous graph neural network HGT \cite{2020-HGT} to encode the dynamics and heterogeneity in each year's heterogeneous network snapshot. The HGT model automatically learns different weights for different types of nodes and relations, and aggregates features accordingly. The node representations learned by HGT serve as dynamic network features before the paper is published.

Since the heterogeneous graph representation learning algorithm is based on the neighbor aggregation mechanism, it takes only the information of a very limited neighborhood for each node. So it can only capture the local evolutionary trend patterns of academic network, unable to capture global trend patterns. As there are huge number of nodes in the academic network interacting with each other, and each node contribute differently, it makes sense to encode the global evolutionary trends to model each node's importance. A paper node with higher importance tends to receive more citations in the following years and vice versa. In node importance embedding module, we take Personalized PageRank (PPR) \cite{1999-ThePageRankCitationRanking} into the graph neural network to reflect the larger neighbor information of different types of relations, capturing the global evolutionary trends. To make full use of the heterogeneity in the graph, we devise a semantic-level attention mechanism to learn the importance of different meta-paths and fuse them automatically. After that, we obtain each new paper's node importance patterns.

In time series generation module, we use attention mechanism to fuse the dynamic node features and node importance patterns learned by above modules to generate final embeddings of each paper. The paper embeddings serve as the parameters of a parametric citation count generator. Based on the work \cite{2013-QuantifyingLongtermScientificImpact}, we use a log-normal distribution to encode prior knowledge of citation processes and generate citation count time series in the years immediately following publication. We unfold the details of these three modules in the following subsections.

\subsection{Dynamic Heterogeneous Network Embedding Module}
\label{section:DynamicHeterogeneousNetworkEmbeddingModule}

A dynamic heterogeneous academic network refers to a sequence of heterogeneous networks from several years before a paper's publication year \emph{T}, and it reflects the evolutionary trends of the entire academic network in different periods. For each year's heterogeneous network, to capture the rich semantic heterogeneity information, we use the Transformer\cite{2017-AttentionIsAllYouNeed}-based heterogeneous graph neural network model HGT \cite{2020-HGT} to learn the node representations of each node, which treats one type of node as query to calculate the importance of other types of nodes around it. As each year's network snapshot reflects different parts of the whole dynamic academic network, and should be comparable, we use the same HGT model to encode each static heterogeneous network in each year. 

Secondly, in academic dynamic network, unlike other dynamic networks such as social dynamic network, nodes won't change rapidly. In other words, the characteristics of same nodes in adjacent years tend to be similar. Inspired by this, we introduce a Mean Squared Error (MSE) loss to force embeddings of same nodes in adjacent years to be close to each other, named temporal-aligned loss:
\begin{equation}
    \label{equ:L_time}
    \mathcal{L}_{time} = \frac{1}{\mathcal{T} - 1} \sum_{t = 1}^{\mathcal{T} - 1} \frac{1}{| V_t \cap V_{t+1} |} \sum_{i \in V_t \cap V_{t+1}} || h_i^t - h_i^{t+1} ||^2_2
\end{equation}
where $h_i^t$ denotes node $i$'s embedding at year $t$ (year starts from 1), $V_t$ denotes the node set in the heterogeneous network of year $t$, and $\mathcal{T}$ denotes the number of years (observed heterogeneous networks) before the paper's publishment. 

By encoding static heterogeneous networks using HGT model, we obtain the embeddings of each node in each year. However, since a newly published paper has no historical citations, it does not exist in the dynamic academic network. As a solution, we use the metadata nodes (e.g. authors, venues, keywords, etc.) existing in previous years which are linked to the new paper to generate the "fake" embeddings of the paper node in past timestamps. The reason is that a paper's metadata nodes are often with a long history and have a high probability to exist in previous years' academic networks. 

To fulfil that, we first explore the snapshot network related to the published year of the paper to find out the linked metadata neighbor nodes, denoted as $N_p$. Then we look back on each previous year's snapshot network and average the features of these neighbor nodes to get the paper's fake embeddings in each past year. As different types of neighbor nodes may not contribute equally to the impact of the paper, inspired by \cite{2021-HINTS}, we apply type-aware trainable weights to preserve the unequal contribution of different kinds of metadata neighbor nodes, as follows:
\begin{equation}
  \label{equ:impute}
  \begin{split}
    \begin{aligned}
    v_p^t = \sum_{r \in \mathbf{R}} \sum_{i \in N_{p,t}^r} W_r \cdot \frac{\mathbf{h}_{i, t}}{| N_{p,t}^m |},
    \end{aligned}
  \end{split}
\end{equation}
where $\mathbf{R}$ denotes the relation set in the network, $\mathbf{h}_{i, t}$ denotes the feature of node $i$ in year $t$, $N_{p, t}^r$ denotes the set of neighbor nodes adjacent to the paper $p$ based on relation $r$, and $W_r$ denotes the learnable weight of relation $r$ shared in all years of network snapshots. After that, the generated fake embeddings of paper $p$ in year $t$, denoted as $v_p^t$, can be obtained.

We apply Eq. \ref{equ:impute} in every timestamp to obtain a sequence of generated fake embeddings of new paper $p$ as $V_p = \{ v_p^t,v_p^{t+1},...,v_p^{T-1} \}$, where $t$ is the first year when paper $p$'s metadata nodes can be observed. Then, to model the new paper $p$'s temporal trajectory, we temporally encode the embedding sequence $V_p$ into a single vector $h_p^g$ through the recurrent neural network LSTM \cite{1997-LSTM} as follows:
\begin{equation}
  \begin{split}
    \begin{aligned}
      h_p^g = \text{LSTM}(v_p^1, v_p^2, ..., v_p^T).
    \end{aligned}
  \end{split}
\end{equation}
After that, we obtain $h_p^g$, the dynamic heterogeneous network feature vector of the paper $p$ before its publication, which reflects new paper's dynamic evolutionary trends. And it will be used in the time series generation module.

\subsection{Node Importance Embedding Module}
\label{sec:NodeImportanceEmbeddingModule}

Since the heterogeneous graph learning algorithm is based on a neighbor aggregation mechanism, it takes only the information of a very limited neighborhood of each node to avoid overfitting and over-smoothing. As a result, the heterogeneous graph neural network can only capture the local consistency relationship of the academic network. However, the global consistency relationship is vitally important. For instance, in an academic network, each scholar can be a member of several communities and can be influenced by his neighborhoods with different distances from local consistency relationship to global consistency relationship, so only considering the local relationship tends to be one-sided. 

To capture the global consistency relationship, we propose a node importance embedding module to calculate each paper's node importance on the global scale. Intuitively, node importance has a close connection with citation counts. A paper with higher node importance has higher academic impacts and tends to receive more citations in the following years.  

In this work, we take Personalized PageRank (PPR) \cite{1999-ThePageRankCitationRanking} into the graph neural network to reflect the global consistency relationship of the whole academic network. We define the PPR matrix as: 
\begin{equation}
  \label{equ:PPR}
  \begin{split}
    \begin{aligned}
      \Pi^{ppr} = \alpha (I_n - (1 - \alpha) D^{-1} A)^{-1},
    \end{aligned}
  \end{split}
\end{equation}
where $\alpha$ is a teleport probability. The PPR representation of node $i$ refers to the $i^{th}$ row in $\Pi^{ppr}$ as $\pi(i) := \Pi^{ppr}_{i,:}$. As the academic network used in our experiment is too large to compute on the whole graph, following the work of \cite{2022-PersonalizedPagerankGraphAttentionNetworks}, we use random walk sampling \cite{2017-InductiveRepresentationLearningOnLargeGraphs} as an efficient and scalable algorithm for computing an approximation of PPR. To guarantee the absolute error lower than $\epsilon$ with probability of $1 - \frac{1}{n}$, we need $O(\frac{\log n}{\epsilon^2})$ random walks. Additionally, we also take the instructions in \cite{2022-PersonalizedPagerankGraphAttentionNetworks} to truncate $\Pi^{ppr}$ to contain only the top $k$ largest entries for each row $\pi(i)$, denoted as $\Pi^{\epsilon, k}_i$.

However, the above method can only handle the homogeneous graph structure. When it comes to heterogeneous graph, since it contains different types of nodes and links, each node is connected via various types of relations, e.g., meta-paths. Since different meta-paths reflect different aspects of the whole graph, and they take unequal contribution to the final result, we compute PageRank respectively in each metapath-based subgraph. Then, inspired by Graph Attention Network (GATv2) \cite{2022-GATv2}, we propose a novel attention mechanism to learn the importance of different meta-paths and fuse multiple semantics revealed by them.

Firstly, we extract metapath-based subgraphs by each meta-path, and compute PageRank matrix $\Pi^{\phi_p}$ for each subgraph using Eq. \ref{equ:PPR}. Then by modifying the attention mechanism proposed in GATv2 \cite{2022-GATv2} to 
\begin{equation}
  \label{equ:ghgat}
  \begin{split}
    \begin{aligned}
      e^{\phi_p}(h_i, h_j) = \alpha^{\phi_p} \cdot \text{LeakyReLU}([W_{\phi_p} h_i || W_{\phi_p} h_j || \Pi^{\phi_p}_i || \Pi^{\phi_p}_j]),
    \end{aligned}
  \end{split}
\end{equation}
where $\alpha^{\phi_p}$ denotes the attention vector of meta-path $\phi_p$, $h_i$ denotes the raw feature vector of node $i$, $W_{\phi_p}$ denotes the transformation weight matrix of meta-path $\phi_p$, aiming at projecting the raw node feature into the meta-path vector space, and $\Pi^{\phi_p}_i$ denotes the PageRank pattern of node $i$ in meta-path $\phi_p$. By Eq. \ref{equ:ghgat}, we can naturally incorporate the global PageRank patterns into the GAT layer in each subgraph.

Then, we normalize the attention scores from all neighbors $j \in \mathcal{N}^{\phi_p}_i$ within meta-path $\phi_p$, and aggregate these features by learned weights:
\begin{equation}
  \begin{split}
    \begin{aligned}
      \alpha^{\phi_p}(h_i, h_j) = \frac{\exp( \text{LeakyReLU}( e^{\phi_p}(h_i, h_j) ) )}{\sum_{k \in \mathcal{N}^{\phi_p}_i} \exp( \text{LeakyReLU}( e^{\phi_p}(h_i, h_k) ) )},
    \end{aligned}
  \end{split}
\end{equation}
\begin{equation}
  \begin{split}
    \begin{aligned}
      z_i^{\phi_p} = \text{LeakyReLU}( \sum_{j \in \mathcal{N}^{\phi_p}_i} \alpha^{\phi_p}(h_i, h_j) \cdot W_{\phi_p} \cdot h_j ).
    \end{aligned}
  \end{split}
\end{equation}
After that, we get node embedding $z_i^{\phi_p}$ for each meta-path $\phi_p$, incorporated with both PageRank patterns and metapath-level representations. Next, in order to fuse these metapath-level representations to get the final node embeddings, we use an attention mechanism similarly, but at the level of meta-path.

Specifically, we do a normalization on each meta-path by averaging the node embeddings $z_i^{\phi_p}$ learned before. Then we use an attention vector $q$ to transform these embeddings into the importance of specific meta-path, denoted as $w^{\phi_p}$:
\begin{equation}
  \begin{split}
    \begin{aligned}
      w^{\phi_p} = q^\text{T} \cdot \frac{1}{|\mathcal{V}|} \sum_{i \in \mathcal{V}} \text{ReLU}(W \cdot z_i^{\phi_p}).
    \end{aligned}
  \end{split}
\end{equation}

To obtain the weight of meta-path $\phi_i$, we normalize the above importance of all meta-paths by softmax function:
\begin{equation}
  \begin{split}
    \begin{aligned}
      \alpha^{\phi_p} = \frac{\exp(w^{\phi_p})}{\sum_{p=1}^P \exp(w^{\phi_p})},
    \end{aligned}
  \end{split}
\end{equation}
where $P$ denotes the number of all meta-paths. Since the learned $\alpha^{\phi_p}$ can be interpreted as the contribution of each meta-path $\phi_p$, we take it as coefficient to fuse semantic-specific embeddings belonging to different meta-paths, to obtain the final embedding as follows:
\begin{equation}
  \begin{split}
    \begin{aligned}
      h^c_i = \sum_{p=1}^P \alpha_{\phi_p} \cdot h^{\phi_p}_i.
    \end{aligned}
  \end{split}
\end{equation}

The final node embedding $h^c_i$ reflects the node importance of paper $i$, capturing the global consistency relationship. It will be used for the time series generation module, together with node features reflecting new paper's dynamic evolutionary trends described in Sec. \ref{section:DynamicHeterogeneousNetworkEmbeddingModule}.

\subsection{Time Series Generation Module}
\label{section:4.4}

Through above two modules, for a target paper $p$, we can generate its dynamic evolutionary trend patterns and node importance patterns in the whole academic network. To combine node dynamic trend patterns and node importance patterns, we use an attention mechanism: 
\begin{equation}
    h_p = \alpha_g h_p^g + \alpha_c h_p^c 
\end{equation}
\begin{equation}
    \alpha_g = \frac{\exp(\lambda^{\text{T}} h_p^g)}{\exp(\lambda^{\text{T}} h_p^g) + \exp(\lambda^{\text{T}} h_p^c)},
\end{equation}
\begin{equation}
    \alpha_c = \frac{\exp(\lambda^{\text{T}} h_p^c)}{\exp(\lambda^{\text{T}} h_p^g) + \exp(\lambda^{\text{T}} h_p^c)},
\end{equation}
where $h_p^g$ refers to the dynamic trend vector of paper $p$ calculated in Sec. \ref{section:DynamicHeterogeneousNetworkEmbeddingModule}, $h_p^c$ refers to the node importance vector of paper $p$ calculated in Sec. \ref{sec:NodeImportanceEmbeddingModule}, and $\lambda$ denotes the trainable attention weight.

As described in \cite{2013-QuantifyingLongtermScientificImpact, 2021-HINTS}, a new paper's influence will reach the peak within a few years after publication, and will gradually decrease on account of novelty fading and continuous appearance of new ideas and new research topics attracting researchers' attention, known as aging effect. Therefore, we model the citation trajectory of a paper as a log-normal probability along time $t$:
\begin{equation}
  \begin{split}
    \begin{aligned}
      P_p(t)=\frac{1}{\sqrt{2\pi}\sigma_pt}exp\left[-\frac{(\ln t - \mu_p)^2}{2\sigma_p^2} \right],
    \end{aligned}
  \end{split}
\end{equation}
where $\mu_p$ denotes the mean of the normal distribution, which describes the time required for an article to reach the peak of citation trajectory. $\sigma_p$ denotes the variance of the normal distribution, which describes the decay rate of paper $p$'s citation decrement.

As discussed in \cite{2021-HINTS}, the "fitness" makes significant contributions to a paper's citations, so another parameter $\eta_p$ is used to model it. Integrated across $\eta_p$, the cumulative number of citations of a paper can be generated by the cumulative distribution function:
\begin{equation}
  \label{equation:12}
  \begin{split}
    \begin{aligned}
      C_p^t = \alpha\left[exp(\eta_p*\Phi(\frac{\ln t-\mu_p}{\sigma_p}))-1 \right],
    \end{aligned}
  \end{split}
\end{equation}
where $\eta_p$ is the parameter which weights the citation count to model the difference between papers. $\alpha$ is a scalar that adjusts the weight of the result, which is a hyper-parameter that will be fixed during the model training process. $\Phi(x)$ is defined as: 
\begin{equation}
  \begin{split}
    \begin{aligned}
      \Phi(x) = (2\pi)^{-1/2}\int_{-\infty}^{x}e^{-y^2/2}dy.
    \end{aligned}
  \end{split}
\end{equation}

To get three parameters $\mu_p$, $\sigma_p$ and $\eta_p$ to generate each new paper's citation time series, we use three Multilayer Perceptron models (MLP) to transform the final node embedding $h_p$ of the target paper $p$ to generate these parameters:
\begin{equation}
  \begin{split}
    \begin{aligned}
      \mu_p = \text{MLP}_1 (h_p),
    \end{aligned}
  \end{split}
\end{equation}
\begin{equation}
  \begin{split}
    \begin{aligned}
      \sigma_p = \text{MLP}_2 (h_p),
    \end{aligned}
  \end{split}
\end{equation}
\begin{equation}
  \begin{split}
    \begin{aligned}
      \eta_p = \text{MLP}_3 (h_p).
    \end{aligned}
  \end{split}
\end{equation}

After that, we can obtain the cumulative citation of target paper $p$ in $1$-$L$ years after publication, denoted as a sequence $C_p = \{ C_p^T, C_p^{T+1}, ..., C_p^{T+L} \}$.

\subsection{Loss Function}

The cumulative citation sequence of target paper $p$ in $1$-$L$ years after publication is calculated by Eq. \ref{equation:12}, denoted as $\{ C_p^T, C_p^{T+1}, ..., C_p^{T+L} \}$. The loss function is composed of two parts: prediction loss and temporal-aligned loss.

The temporal-aligned loss is discussed in Sec. \ref{section:DynamicHeterogeneousNetworkEmbeddingModule}. As nodes in academic dynamic network won't change rapidly and the characteristics of same nodes in adjacent years tend to be similar, the temporal-aligned loss aims to force embeddings of same nodes in adjacent years to be close to each other, as described in Eq. \ref{equ:L_time}.

For prediction loss, we adopt the Mean Square Error (MSE) to compare the predicted time series with the ground-truth as prediction loss:
\begin{equation}
  \label{equ:PredictionLoss}
  \begin{split}
    \begin{aligned}
      \mathcal{L}_{pred} = \frac{1}{P}\sum_{p=1}^P \frac{1}{L}\sum_{t=T+1}^{T+L}(C_p^{t} - \hat{c}_p^t)^2,
    \end{aligned}
  \end{split}
\end{equation}
where $C_p^t$ denotes the ground-truth citation count of paper $p$ in the $t^{th}$ year, $T$ denotes the number of prediction years, and $P$ denotes the total number of papers for prediction. Since the citation counts of different papers vary widely, we conduct log transformation on ground-truth citation counts to smooth rapid changes, i.e. $c_p^{\hat{t}} = \log(C_p^{\hat{t}} + 1)$, and make predictions on the logged version of true values. 

The overall model will be optimized by prediction loss (Eq. \ref{equ:PredictionLoss}) and temporal-aligned loss (Eq. \ref{equ:L_time}) at the same time. The total loss is defined as follow:
\begin{equation}
    \mathcal{L} = \mathcal{L}_{pred} + \beta \mathcal{L}_{time}
\end{equation}
where $\beta$ is the hyper-parameter to adjust the proportion of temporal-aligned loss in total loss.

\section{Experiments}

In this section, we evaluate the performance of our proposed model DGNI by experiments on two real-world large-scale datasets. We describe our experimental settings and then show numerical comparison results with other citation count prediction baselines. To help readers understand how DGNI works, we breakdown the model in ablation studies and conduct visualization analyses to prove DGNI's efficacy.

\subsection{Experimental Setup}

\subsubsection{Datasets}

We conduct experiments on two real-world datasets \emph{APS} and \emph{AMiner}. The statistics of nodes and edges about the two datasets are shown in Table \ref{tab:dataset}. Below we take a brief introduction:

\textbf{APS}\footnote{\url{https://journals.aps.org/datasets}} (\emph{American Physical Society}) is a dataset that covers publications in the journal of the American Physical Society, including three node types: paper, author, venue. To generate keyword nodes, we extract keywords from the title of papers following the pre-processing procedure proposed by \cite{2018-AutomatedPhraseMiningFromMassiveTextCorpora}. In the experiment, We use papers from 2003 to 2008 to build the training set to train the model, papers in 2009 to build the validation set, and papers in 2010 to build the testing set.

\textbf{AMiner}\footnote{\url{https://aminer.org/citation}} is a dataset that covers publications in computer science venues \cite{tang2008arnetminer}, including four node types: paper, author, venue and keyword in its V11 version. The training, validation and testing datasets are of the same configuration as APS.

\begin{figure}[ht]
\centering
    \centering
    \vspace{-0.3cm}
    \includegraphics[width=1.0\columnwidth]{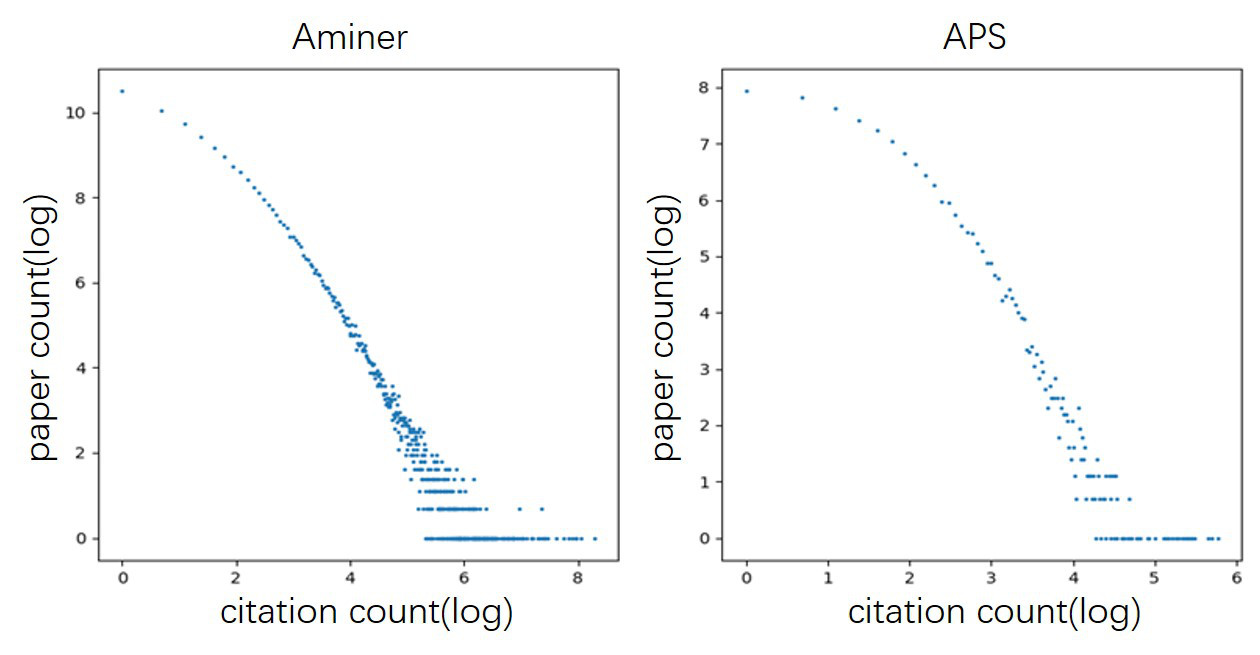}
    \vspace{-0.3cm}
    \caption{Distribution of cumulative citation counts within five years after publication.}
    \label{fig:longtail}
    \vspace{-0.3cm}
\end{figure}

The distribution of cumulative citation counts of papers in two datasets is shown in Fig. \ref{fig:longtail}. We can note that the cumulative citations of papers subject to the long-tailed distribution: most papers are rarely cited after publication, and only a fraction of them can receive considerable citations.

\begin{table}
  \newcommand{\tabincell}[2]{\begin{tabular}{@{}#1@{}}#2\end{tabular}}
  \centering
  \caption{The Statistics of Datasets.}
   \vspace{-0.2cm}
  \label{tab:dataset}
  \resizebox{1.0\columnwidth}{!}{
    \begin{tabular}{c|cccc|c}
      \toprule
      \multirow{2}{*}{Dataset} & \multicolumn{4}{c|}{\#node} & \multirow{2}{*}{\#edge} \\
      & \#paper & \#author & \#keyword & \#venue \\
      \midrule
      APS & 311,533 & 161,051 & 41,126 & 9 & 3,250,651 \\
      AMiner & 1,026,795 & 831,151 & 34,833 & 3,673 & 10,366,576 \\
      \bottomrule
    \end{tabular}
  }
  \vspace{-0.5cm}
\end{table}

\subsubsection{Baselines}\label{para:5.2}

Since the "cold start" citation count time series prediction task is a novel problem, there is only one work (i.e. HINTS \cite{2021-HINTS}) to compare with. Besides, we consider 4 other citation count time series prediction methods for comparison, which are briefly described below.

\begin{itemize}
\item \textbf{Gradient Boosting Machine (GBM):} A gradient boosting model used to model scientific features and predict citation time series. Following \cite{2021-HINTS}, we extract scientific features that are available in our problem setting or data, to predict citation time series with XGBoost \cite{chen2016xgboost}.

\item \textbf{DeepCas \cite{2017-Deepcas}:} This model conducts random walk across an information cascade graph to predict popularity. In our experiments, the ego network of a new paper in the publication year is used as the initial cascade graph.

\item \textbf{HINTS \cite{2021-HINTS}:} A state-of-the-art model for "cold start" citation count time series prediction. This model uses R-GCN \cite{2018-RGCN} to encode dynamic heterogeneous graph and a log-normal distribution to generate the citation time series.

\end{itemize}

\subsubsection{Evaluation Metrics}
Following \cite{2017-Deepcas, 2017-DeepHawkes, 2021-HINTS}, we use the Mean Absolute Error (MAE) and Root Mean Squared Error (RMSE) to evaluate the accuracy of predictions, which are common choices for regression tasks. MAE and RMSE are defined as follows:
\begin{equation}
  \label{equ:MAE}
  \text{MAE}(c^t, c^{\hat{t}}) = \frac{1}{P}\sum_{p=1}^P| c_p^{\hat{t}} - c_p^t |,
\end{equation}
\begin{equation}
  \label{equ:RMSE}
  \text{RMSE}(c^t, c^{\hat{t}}) = \sqrt{\frac{1}{P}\sum_{p=1}^P (c_p^{\hat{t}} - c_p^t)^2},
\end{equation}
where $c^t_p$ and $c_p^{\hat{t}}$ denote the ground-truth and the prediction of citation counts of paper $p$ in the $t^{th}$ year after publication, and $P$ denotes the total number of papers for prediction.

\subsubsection{Implementation Details}

We implement DGNI using PyTorch 1.10.0. For the dynamic heterogeneous network embedding module, the number of historical dynamic heterogeneous networks we use to model is 3, the dimension of output features of HGT is 32, the layer count and attention heads are set to 2 and 4. The node features in dynamic heterogeneous network are randomly initialized using Xavier initialization. The hidden dimension and the layer count of GRU are 32 and 3 respectively. 

For node importance embedding module, the number of attention heads and attention layers are 4 and 2 respectively, the hidden dimension is set to 32. The meta-paths we use are based on paper nodes: PAP (Paper-Author-Paper), PVP (Paper-Venue-Paper), PKP (Paper-Keyword-Paper). For time series generation module, the hidden dimensions of three fully-connected layers are all set to 20. The weight of temporal-aligned loss $\beta$ is set to 0.5. 

We set learning rate to 0.001 and model parameter optimizer as Adam. We set batch size to 3000 for both APS and AMiner datasets. All the models predict the citation series of the paper in the first 5 years after publication, and the averages are used as the prediction result. We run each experiment 10 times following the same configuration with different random seeds and take the average of all results as final result.

\renewcommand\arraystretch{1.2}

\begin{table*}
  \newcommand{\tabincell}[2]{\begin{tabular}{@{}#1@{}}#2\end{tabular}}
  \centering
  \caption{Comparison Results of Different Methods over Two Datasets.}
  \vspace{-0.2cm}
  \label{tab:result2}
  \resizebox{1.8\columnwidth}{!}{
  \begin{tabular}{c|c|cccccc|cccccc}
  \toprule
  \multirow{2}{*}{Dataset} &
  \multirow{2}{*}{Model} &
  \multicolumn{6}{c|}{MAE} &
  \multicolumn{6}{c}{RMSE} \\
  \cline{3-14}
  & & year1 & year2 & year3 & year4 & year5 & overall & year1 & year2 & year3 & year4 & year5 & overall  \\
  \midrule

  \multirow{4}{*}{APS}
  & GBM & 0.898 & 0.885 & 0.900 & 0.921 & 1.041 & 0.934 & 1.098 & 1.088 & 1.105 & 1.124 & 1.274 & 1.139\\
  & DeepCas & 0.931 & 0.923 & 0.885 & 0.855 & 0.832 & 0.904 & 1.125 & 1.139 & 1.126 & 1.103 & 1.062 & 1.104\\
  & HINTS & 0.769 & 0.809 & 0.825 & 0.831 & 0.828 & 0.805 & 0.936 & 0.994 & 1.019 & 1.032 & 1.035 & 1.023\\
  \cline{2-14}
  & DGNI & \textbf{0.608} & \textbf{0.689} & \textbf{0.734} & \textbf{0.766} & \textbf{0.788} & \textbf{0.717} &
  \textbf{0.748} & \textbf{0.850} & \textbf{0.909} & \textbf{0.949} & \textbf{0.978} & \textbf{0.891}\\
  \midrule

  \multirow{4}{*}{AMiner}
  & GBM & 0.584 & 0.920 & 0.989 & 1.310 & 1.260 & 1.018 & 0.691 & 1.031 & 1.224 & 1.535 & 1.621 & 1.224\\
  & DeepCas & 0.948 & 1.052 & 1.008 & 0.898 & 0.968 & 0.981 & 1.054 & 1.260 & 1.302 & 1.245 & 1.265 & 1.258\\
  & HINTS & 0.610 & 0.710 & 0.751 & 0.775 & 0.788 & 0.764 & 0.769 & 0.905 & 0.966 & 1.001 & 1.024 & 0.991\\
  \cline{2-14}
  & DGNI & \textbf{0.491} & \textbf{0.629} & \textbf{0.704} & \textbf{0.759} & \textbf{0.803} & \textbf{0.677} & \textbf{0.606} & \textbf{0.782} & \textbf{0.898} & \textbf{0.986} & \textbf{1.003} & \textbf{0.879} \\
  \bottomrule
  \end{tabular}
  }
\end{table*}

\subsection{Numerical Comparison Results}

\subsubsection{Comparison with Baselines}

The prediction results are shown in Table \ref{tab:result2}. We can summarize that DGNI achieves significant improvement on all the baselines on two datasets in terms of both MAE and RMSE. Since HINTS is the baseline with best performance as we know, for our proposed model DGNI, in terms of MAE, DGNI outperforms HINTS by 10.93\% on APS and 11.39\% on AMiner. And for RMSE, DGNI outperforms HINTS by 12.90\% on APS and 11.31\% on AMiner. The results prove the effectiveness and efficacy of our DGNI. 

Compared with HINTS, our use of HGT as the dynamic heterogeneous network encoder has stronger feature expression ability than the use of simple R-GCN. Besides, our proposed node importance embedding module can capture the node importance of different papers and pay more attention on papers with higher reputation, thus boosting the prediction of citation count. The more detailed analyses of these two components are elaborated in Ablation Study in Sec. \ref{sec:ablation}.

Additionally, it can be found from annual prediction results that DGNI can achieve the best results also in the annual results and can achieve better performance in early citation prediction than long-term citation prediction.

\subsubsection{Ablation Study on DGNI Components}
\label{sec:ablation}

\begin{table}
  \newcommand{\tabincell}[2]{\begin{tabular}{@{}#1@{}}#2\end{tabular}}
  \centering
  \caption{The Ablation Results of DGNI on Two Datasets.}
  \vspace{-0.2cm}
  \label{tab:ablation}
  \resizebox{0.8\columnwidth}{!}{
    \begin{tabular}{c|c|c c c c c|c}
    \toprule
    Dataset & Metric & \tabincell{c}{DGNI-graph} & \tabincell{c}{DGNI-inp} & DGNI \\
    \midrule
    \multirow{2}{*}{APS} & MAE & 0.728 & 0.805 & \textbf{0.717}\\
    & RMSE & 0.909 & 0.987 & \textbf{0.891}\\
    \midrule
    \multirow{2}{*}{AMiner} & MAE & 0.699 & 0.684 & \textbf{0.677}\\
    & RMSE & 0.914 & \textbf{0.866} & 0.879\\
    \bottomrule
    \end{tabular}
  }
\end{table}

To present more detailed analyses of these components in our proposed DGNI model and find out why DGNI works, we compare DGNI with two variants on APS and AMiner datasets to evaluate the effectiveness of the modules of our model. The two variants are described as follows: 
\begin{itemize}
    \item \textbf{DGNI-graph}: A variant of our framework, which removes the node importance embedding module and only uses dynamic heterogeneous network for prediction.
    
    \item \textbf{DGNI-inp}: A variant of our framework, which removes the dynamic heterogeneous network embedding module and only uses node importance embedding module for prediction.
\end{itemize}

The experimental results are shown in Table \ref{tab:ablation}.
From the results, we can come to the following conclusions: (1) The whole DGNI can achieve almost best results than all variants on APS and AMiner datasets. It verifies the effectiveness of all our proposed modules in DGNI. 
(2) The node importance embedding module has a great impact on the results, because after removing the module, the model accuracy on both datasets gets lower. The reason might be that the node importance can reflect the global consistency relationship of the network and can guide the allocation of focus on different papers. Furthermore, in terms of AMiner dataset, the variant of DGNI-inp has the best performance, the reason might be that in AMiner dataset the global consistency relationship is more important.
(3) The use of dynamic heterogeneous network plays a key role in the citation prediction. As after removing the dynamic heterogeneous network embedding module, the model performance on both datasets has a huge decrease. The reason might be that heterogeneous graph neural model can capture the rich structure information in the network. And structure information is the core of heterogeneous academic network and vital for citation prediction.

\subsection{Visualization Analysis}

\subsubsection{Feature Dimension Reduction Analysis}

\begin{figure}[htbp]
\centering
    \centering
    \includegraphics[width=0.6\columnwidth]{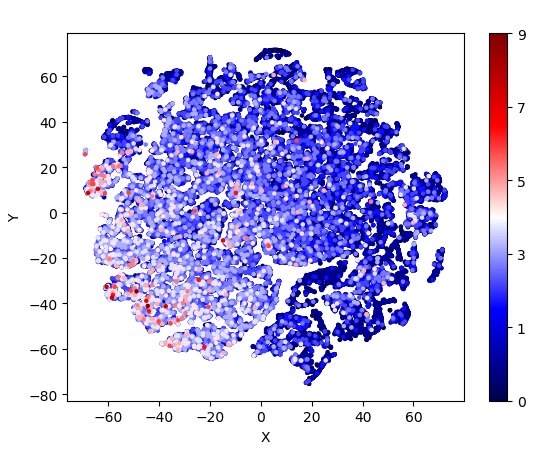}
    \vspace{-0.3cm}
    \caption{The T-SNE projection result of final embeddings on AMiner dataset. Each point represent the log-scale cumulative citation count.}
    \label{fig:low}
    \vspace{-0.5cm}
\end{figure}

In order to verify that DGNI can learn expressive embeddings for each paper, we use T-SNE \cite{2008-tSNE} to project final embeddings $h_p$ on AMiner dataset into a two-dimensional space, as shown in Fig. \ref{fig:low}.

The figure represent the log-scale cumulative citation count in 5 years after publication. The blue points indicate low-cited papers, while red points indicate high-cited papers. It can be seen that the citation counts from the red points (left-bottom) to the blue points (right-top) decrease gradually, so DGNI can model papers with different citation counts effectively. But there are still many lowly cited papers and high-cited papers distribute together. The reason is that the specific number of citations of the paper are more related to the quality of the paper itself. Therefore, DGNI is more discriminative on the macro scale.

\subsubsection{Normal Distribution Parameter Analysis} 

\begin{figure}[ht]
  \centering
      \centering
      \includegraphics[width=0.6\columnwidth]{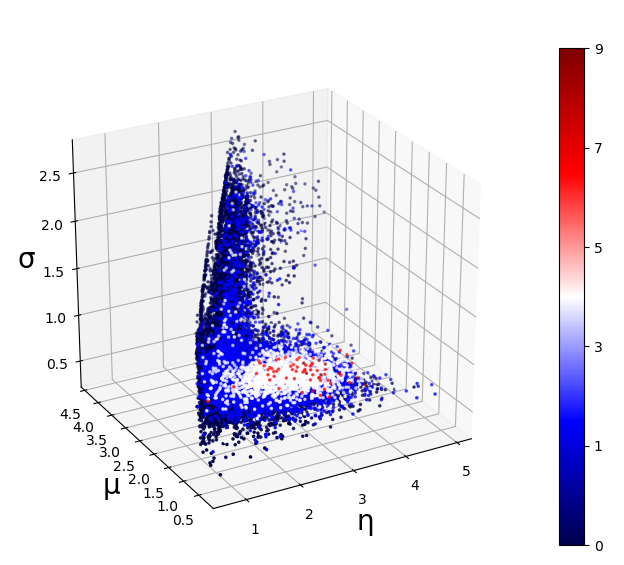}
      \vspace{-0.3cm}
      \caption{The visualization of parameters. Each point represent the log-scale cumulative citation count.}
      \label{fig:3d}
      \vspace{-0.3cm}
  \end{figure}

Then, we visualize the relationship between the parameters of the normal distribution and the cumulative number of citations, the result is shown as Fig. \ref{fig:3d}. The cumulative citation counts gradually increase from the left to the right. At the same time, the more the citation counts, the larger the parameter of the normal distribution $\eta_p$. In addition, it can be seen that the high-cited papers usually have smaller parameters $\sigma$, which indicates that high-cited papers have a larger weight and higher growth rate. The conclusion is in line with the physical meaning of the parameters we proposed in Sec. \ref{section:4.4}.

\begin{figure*}[ht]
  \centering
      \centering
      \vspace{-0.2cm}
      \includegraphics[width=1.4\columnwidth]{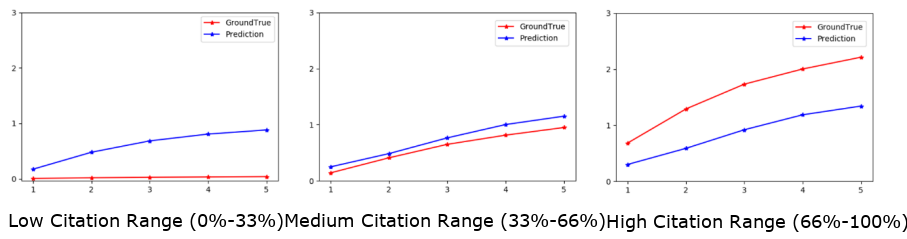}
      \vspace{-0.3cm}
      \caption{The predicted citation counts compared with ground-truth of papers with different citation range.}
      \label{fig:PredictionErrorAnalysis}
      \vspace{-0.5cm}
\end{figure*}

\subsubsection{Prediction Error Analysis}

In order to analyze the prediction error between papers of different citation counts, we experiment on AMiner dataset and compare the gap between actual value and predicted value. Specifically, we evenly divide the dataset into three parts by papers' citation numbers: low citation number interval (0\%-33\%), medium citation number interval (33\%-66\%) and high citation number interval (66\%-100\%). For visualization analysis, we plot the average of prediction result and actual result in the following 5 years after publication, on the line chart.

As shown in Fig. \ref{fig:PredictionErrorAnalysis}, the DGNI model has a better prediction accuracy in the medium citation count range, but a higher prediction error in the low citation count range and high citation count range. The reason might be the long-tailed distribution of paper citations. Exactly, most papers included in the real-world dataset can only receive 0 or 1 citations in the following 5 years after publication, while only a small part of papers can receive higher citation numbers. However, the low citation result is not necessarily depended on authors, venues or fields, since a prominent scholar may also publish papers without citations, and papers in top-ranked journals or conferences often receive no citations either. As a result, on the one hand, the independence between papers and their related metadata brings challenges to the prediction of papers with low citation counts. On the other hand, since papers in high citation count range only constitute a small part of the whole dataset, leading to the poor prediction performance in such a scenario.

\section{Related Work}
This section reviews three lines of related work: citation time series prediction, heterogeneous graph representation learning, and node importance estimation.

\subsection{Citation Time Series Prediction}
Citation count prediction includes two categories: using early citation after publication to predict and using information before publication to predict. Parametric approaches uses early citations to model citation trends as a parametric pattern \cite{2014-ModelingAndPredictingPopularityDynamicsViaReinforcedPoissonProcesses,ke2015defining,2016-OnModelingAndPredictingIndividualPaperCitationCountOverTime,2017-OnPredictivePatentValuation}. Some researchers use machine learning methods to model the early citations and citation graph after publication, e.g. Abrishami and Aliakbary \cite{2019-PredictingCitationCountsBasedOnDeepNeuralNetworkLearningTechniques} used the schema of encoder-decoder to convert the early citations into future citation trends; Li et al. \cite{2017-Deepcas} modeled the early citations of the paper as an information cascade network. There are similar methods like \cite{2017-UnderstandingTheImpactOfEarlyCitersOnLongtermScientificImpact,2020-UtilizingCitationNetworkStructureToPredictCitationCounts}. However, this kind of method relies on the early citation within 1-3 years after publication, so it can not deal with the cold start problem.
 
Some recent works focus on predicting future citations of new papers \cite{2011-CitationCountPrediction,2016-CanScientificImpactBePredicted}. Li et al. \cite{2019-ANeuralCitationCountPredictionModelBasedOnPeerReviewText} used peer review text to predict future citations. Xu et al. \cite{2019-EarlyPredictionOfScientificImpactBasedOnMultibibliographicFeaturesAndConvolutionalNeuralNetwork} used heterogeneous academic network to predict the citations of the paper in ten years. The work \cite{2021-HINTS} is the first work to generate citation time series for newly-published paper without any leading citation values, and they use dynamic GNN to model the dynamic academic networks before the publication of papers. Despite most of the citation prediction of newly-published papers are proposed, they are not using information effectively. In this paper, we propose a new framework, which uses academic networks and node importance to predict citation of newly-published papers.

\subsection{Node Importance Estimation}

Early node importance estimation methods used the degree of nodes in the networks to measure the importance of nodes \cite{2002-TopicSensitivePageRank,2008-RandomWalkWithRestart}. PageRank \cite{1999-ThePageRankCitationRanking} is a classic algorithm based on random walk model to propagate the importance of each node to another node with a certain probability. By traversing the entire graph, the importance of each node can be quickly calculated. With the rapid development of deep learning in recent years, some methods based on graph neural networks are proposed. Park et al. \cite{2019-EstimatingNodeImportanceInKnowledgeGraphsUsingGNN} estimated node importance by using graph attention mechanism, and fusing information of neighbor nodes. Huang et al. \cite{2021-RepresentationLearningOnKnowledgeGraphsForNodeImportanceEstimation} utilized a relational graph transformer to learn semantic features, and node2vec \cite{2016-Node2Vec} to learn structure features. Then combine these features to get the final node importance values. However, these methods can only predict the node importance at a single time rather than time series. In addition, they cannot model the evolution of dynamic network.

\subsection{Heterogeneous Graph Representation Learning}

Recent years have witnessed the emerging success of graph neural networks (GNNs) for modeling graph structured data \cite{2019-AComprehensiveSurveyOnGNN}. While most GNNs only work for homogeneous graphs, to represent heterogeneous structures and capture the dynamics of network time series, which are more associated with real-world scenarios, researchers have further developed heterogeneous GNNs \cite{2020-ASurveyOnHeterogeneousGraphEmbedding} and dynamic GNNs \cite{2020-RepresentationLearningForDynamicGraphsASurvey}. For example, Schlichtkrull et al. \cite{2018-RGCN} propose RGCN using multiple weight matrices to project the node embeddings into different relation spaces to capture the heterogeneity of the graph. Wang et al. \cite{2019-HAN} propose HAN using a hierarchical attention mechanism to capture both node and semantic importance. Hu et al. \cite{2020-HGT} propose HGT treating one type of node as query to calculate the importance of other types of nodes around it, by multi-head attention mechanism.

\section{Conclusion}

In this paper, we develop a framework named DGNI which adaptively fuses the dynamic evolutionary trends and the node importance in the academic network to predict citation time series with parameter-based generator, tackling the problem of cold start citation count prediction. The contrast experiments and ablation experiments have been conducted on two real-world datasets to demonstrate the superiority of our framework and effectiveness of all the components respectively. For future work, we will consider the interaction between the importance of nodes and more effective generator.

\begin{acks}
This research was supported by by National Key R\&D Program of China under Grant 2019YFA0707204 and the National Natural Science Foundation of China under Grant Nos. 62176014.
\end{acks}

\bibliographystyle{ACM-Reference-Format}
\balance
\bibliography{sample-base}

\end{document}